# A Measurement of the Branching Ratio and Form Factor of $K_L \to \mu^+\mu^-\gamma$

M. B. Spencer, K. Arisaka, D. Roberts*, W. Slater, M. Weaver

*University of California at Los Angeles, Los Angeles, California 90024*

R. A. Briere[†], E. Cheu, D. A. Harris[†], P. Krolak, K. S. McFarland[‡], A. Roodman,

B. Schwingenheuer, S. V. Somalwar[§], Y. W. Wah, B. Winstein, R. Winston

*The Enrico Fermi Institute, The University of Chicago, Chicago, Illinois 60637*

A. R. Barker

*University of Colorado, Boulder, Colorado 80309*

E. C. Swallow

*Elmhurst College, Elmhurst, Illinois 60126,*

*and The Enrico Fermi Institute, The University of Chicago, Chicago, Illinois 60637*

G. J. Bock, R. Coleman, M. Crisler, J. Enagonio[‖], R. Ford, Y. B. Hsiung,

D. A. Jensen, E. Ramberg[¶], R. Tschirhart

*Fermi National Accelerator Laboratory, Batavia, Illinois 60510*

E. M. Collins**, G. D. Gollin

*University of Illinois, Urbana, Illinois 61801*

T. Nakaya, T. Yamanaka

*Department of Physics, Osaka University, Toyonaka, Osaka, 560 Japan*

P. Gu, P. Haas[††], W. P. Hogan, S. K. Kim[‡‡], J. N. Matthews, S. S. Myung[‡‡],

S. Schnetzer, G. B. Thomson, Y. Zou

*Rutgers University, Piscataway, New Jersey 08855*

## Abstract

We report on a measurement of the decay $K_L \to \mu^+\mu^-\gamma$ from Fermilab experiment E799. We observe 207 candidate signal events with an estimated background of $10.5 \pm 4.0$ events and establish $B(K_L \to \mu^+\mu^-\gamma) = (3.23 \pm 0.23(stat) \pm 0.19(sys)) \times 10^{-7}$. This provides the first measurement of the $K\gamma\gamma^*$ form factor in the muonic Dalitz decay mode of the $K_L$.

Accepted for publication in Physical Review Letters.

The decay $K_L \to \mu^+\mu^-\gamma$ is expected to be dominated by the so-called long distance contributions occurring via the $K\gamma\gamma^*$ vertex [1]. Knowledge of long distance contributions can help to determine the origin of the $\Delta I = \frac{1}{2}$ enhancement in non-leptonic weak interactions. Other rare kaon decays such as $K_L \to \mu^+\mu^-$ depend on similar long distance contributions which must be measured in order to extract standard model parameters [2–4]. The first study of the decay $K_L \to \mu^+\mu^-\gamma$ is an important test of the current theoretical understanding of long distance effects in neutral kaon decays.

The differential decay rate for $K_L \to l^+l^-\gamma$ can be calculated using QED modified by a form factor $f(x)$ for the $K\gamma\gamma^*$ vertex. The result normalized to $K_L \to \gamma\gamma$ is given by

$$\Gamma_{\gamma\gamma}^{-1}\frac{d\Gamma}{dx}(K_L \to l^+l^-\gamma) = \frac{2\alpha}{3\pi}\frac{|f(x)|^2}{x} \times$$
$$(1-x)^3\left(1+\frac{2m_l^2}{xm_K^2}\right)\left(1-\frac{4m_l^2}{xm_K^2}\right)^{1/2} \quad (1)$$

where $x = m_{ll}^2/m_K^2$, and the normalization is such that $f(0) = 1$. Bergström, Massó, and Singer (BMS) [1] parametrized this form factor using a combination of a vector meson dominance model ($K_L \to \pi^0, \eta, \eta' \to \gamma^*\gamma$) and a vector-vector transition model ($K_L \to K^*\gamma \to (\rho,\omega,\phi)^*\gamma \to \gamma^*\gamma$). In this model there is a single free parameter $\alpha_{K^*}$ and $f(x) = f_1(x) + \alpha_{K^*}f_K(x)$, where $f_1(x)$ represents the vector meson dominance amplitudes and $f_K$ is a sum over pole terms from a model of the $KK^*\gamma$ vertex. The parameter $\alpha_{K^*}$ has been inferred from studies of the decay mode $K_L \to e^+e^-\gamma$ [5,6]. The model of Ko [4] has no free parameters, and makes a definite prediction for the form factor. At this point chiral perturbation theory gives no prediction for $K_L \to \mu^+\mu^-\gamma$ as diagrams cancel at $\mathcal{O}(p^4)$ and higher order terms have not been calculated [7].

In the decay $K_L \to l^+l^-\gamma$ the di-lepton invariant mass distribution is constrained to be between $2m_l$ and $m_K$. Together with the $1/x$ dependence of the differential rate in Eq. 1 this gives the electron mode a much larger region of decay phase space as compared to the muon mode. Thus measurements of the form factor $f(x)$ will be most sensitive to the di-electron mass spectrum (and not the $K_L \to e^+e^-\gamma$ branching fraction) and the $K_L \to \mu^+\mu^-\gamma$ branching fraction (but not the $K_L \to \mu^+\mu^-\gamma$ di-muon mass spectrum) assuming uniform experimental acceptances.

Fermilab experiment E799 was designed to search for rare CP violating $K_L$ decays including $K_L \to \pi^0 e^+ e^-$ and $K_L \to \pi^0 \mu^+ \mu^-$ [8]. In the fixed target mode of the Tevatron, 800 $GeV$ protons were incident on a beryllium target. Two nearly parallel neutral beams were produced including $K_L$'s which then decayed in flight. The detector is described in detail elsewhere [9,10]; only those elements relevant to this analysis are mentioned here. Two hodoscope banks of vertical and horizontal counters 175 $m$ downstream of the target were used to trigger on kaon decays with two or more charged tracks. Just downstream of these was a circular calorimeter of 804 lead-glass blocks with cross sections $5.8 cm \times 5.8 cm$ and 18.7 radiation lengths long. The resolution for typical photon energies (12.5 $GeV$) in the calorimeter was 5%. Four drift chambers each with single hit resolutions of 100 $\mu m$ and a magnet with a transverse momentum kick of $200 MeV/c$ formed a spectrometer that measured charged track momenta with resolution $(\sigma_p/p)^2 = (5 \times 10^{-3})^2 + (1.4 \times 10^{-4} p/(GeV/c))^2$. In-time energy deposits in the calorimeter were identified with a hardware cluster finder, and a fast energy sum for the entire calorimeter was used for triggering. In addition there were several annular photon veto counters surrounding the vacuum decay volume which enabled the rejection of events with decay products outside the acceptance region of the calorimeter. Just behind the calorimeter was a 10 cm thick lead wall followed by a hodoscope bank of 45 overlapping counters designed to veto hadronic showers initiating in the calorimeter. The hadron veto was determined by an analog sum of all the individual signals from the counters. Following the hadron veto counter was a muon filter comprised of 3 $m$ (20 interaction lengths) of steel. Muons traversing the filter steel were identified by a hodoscope bank consisting of 16 vertical, non-overlapping counters.

In order to trigger on events with two muons we required the following elements: at least two hits in each of the trigger hodoscope banks; no hits in the photon veto counters; at least two non-adjacent hits in the muon-hodoscope bank; a sufficient number of hits in the drift chambers to be consistent with two tracks; no activity in the hadron veto counter; and a minimum energy sum of 6 $GeV$ in the calorimeter. A two-track "minimum-bias" trigger simultaneously collected $K_L \to \pi^+\pi^-\pi^0$ events which were used for normalization. This trigger was identical to the dimuon trigger but had no muon hodoscope, hadron hodoscope or minimum energy deposit requirements. The $K_L \to \pi^+\pi^-\pi^0$ trigger was prescaled by a factor of 3600 relative to the $K_L \to \mu^+\mu^-\gamma$ trigger.

Data were collected in the 1991 fixed target run during which 60 million dimuon triggers were recorded. Data reduction began by selecting events which had two good reconstructed tracks that projected to clusters in the calorimeter and one "photon cluster" not associated with tracks. These events were almost entirely $K_L \to \pi^\pm \mu^\mp \nu(\gamma_{acc})$ or $K_L \to \pi^+\pi^-\pi^0$ decays. In the first case the pion is misidentified as a muon or the pion decayed, and $\gamma_{acc}$ is an "accidental" photon, i.e. not associated with the $K_L$ decay. For $K_L \to \pi^+\pi^-\pi^0$ decays, both pions were misidentified as muons, and one of the photons from the decay of the $\pi^0$ escaped or overlapped a track cluster. The photon cluster was required to match a cluster found by the hardware cluster finder which had $30 ns$ charge integration gate as opposed to the $100 ns$ used to measure the energy deposited in each block. This suppressed events with out-of-time energy deposits. To further suppress contamination from accidental deposits, the photon cluster centroids were required to be at least 0.5 blocks from the beam holes in the calorimeter. Events with photon cluster energies below 8 $GeV$ were rejected since accidental photons in background events typically have lower energy. This reduced the signal by 9% and rejected 42% of $K_L \to \pi^\pm \mu^\mp \nu(\gamma_{acc})$ events. Clusters not associated with tracks were required to have transverse profiles consistent with the shape expected from a photon. Track momenta were required to be above 7 $GeV/c$, greater than the 5 $GeV/c$ threshold for muons to pass through 3 $m$ of steel. Tracks with associated energy deposits above 3 $GeV$ in the calorimeter were rejected, thus suppressing events with hadron showers in the calorimeter. Monte Carlo studies of $K_L \to \mu^+\mu^-\gamma$ events show that the photon clusters in background events are frequently near the beamholes in the calorimeter, and are therefore on average closer to tracks. We use this correlation to reject background events by requiring the separation of the photon cluster and the tracks at the calorimeter to be greater than 20 $cm$. This cut reduces the signal by 11% and the $K_L \to \pi^\pm \mu^\mp \nu(\gamma_{acc})$ background by 33%.

For events passing these cuts, the transverse momentum squared ($P_t^2$) versus the reconstructed invariant $\mu^+\mu^-\gamma$ mass is shown in Fig. 1. There is a clear signal in the region near the $K_L$ mass and low $P_t^2$. The invariant mass distribution for events with $P_t^2 < 200 MeV^2/c^2$ is shown in Fig. 2. The region $P_t^2 < 200 MeV^2/c^2$ and $\mu^+\mu^-\gamma$ invariant mass in the range $0.482 < M_{\mu\mu\gamma} < 0.514\ GeV/c^2$ was chosen so as to optimize the statistical sensitivity to the branching ratio. For the Monte Carlo simulated data that passed the previous cuts, this signal box contains 77% of the $K_L \to \mu^+\mu^-\gamma$ events, and less than 0.5% of the background events.

The large band at low $\mu^+\mu^-\gamma$ mass in Fig. 1 is consistent with $K_L \to \pi^+\pi^-\pi^0$ decays, and it falls sharply with increasing mass, producing a negligible contribution under the signal peak. We also considered the radiative process $K_L \to \pi^\pm \mu^\mp \nu \gamma$ as a potential source of background, however, it also falls rapidly with increasing mass and has negligible contribution at the kaon mass. The $K_L \to \pi^\pm \mu^\mp \nu (\gamma_{acc})$ background events produce a broad distribution in $\mu^+\mu^-\gamma$ invariant mass and all of the background under the signal peak is consistent with having come from this process. All of the background processes considered have uniform $P_t^2$ distributions, and we determined their combined mass-shape by using a region of data at $400 < P_t^2 < 1000\ MeV^2/c^2$ (Fig.2). The total background is estimated to be $10.5 \pm 4.0$ events out of 207 candidate signal events.

A similar analysis was performed on the minimum bias data to reconstruct $K_L \to \pi^+\pi^-\pi^0$ decays for normalization. The major difference is that now two photon clusters were required, and there was no 3 $GeV$ track-cluster energy cut. The cut on the photon energy was relaxed to 7 $GeV$, thus increasing the yield. In addition, a cut of $3\sigma$ (21 $MeV$) around the $\pi^0$ mass was made on the reconstructed invariant mass of the two photon clusters.

High statistics Monte Carlo samples were generated for the normalization and signal modes. To properly account for activity in the detector not associated with the parent kaon decay, special "accidental" events were collected in parallel with the dimuon and minimum bias triggers. These accidental events sample the instantaneous rate of activity in the detector throughout the run, and are overlaid on the simulated decays. This simulated data were then passed through an identical analysis to that described above. Acceptances calculated by Monte Carlo for kaons decaying in the momentum range of 20-220 $GeV/c$ and in a region 90-160 m from the target were 1.95% for $K_L \to \mu^+\mu^-\gamma$ and 1.95% for $K_L \to \pi^+\pi^-\pi^0$. With 196.5 signal events and a final sample of 20,919 $K_L \to \pi^+\pi^-\pi^0$ events, the branching ratio is $B(K_L \to \mu^+\mu^-\gamma)/B(K_L \to \pi^+\pi^-\pi^0) = (2.61 \pm 0.19(stat)) \times 10^{-6}$.

The largest sources of systematic error in this ratio are due to differences between the dimuon and minimum bias triggers, the dominant contribution being from the hadron shower veto. Special muon calibration runs as well as minimum bias data were used to determine the absolute dimuon efficiency of the hadron veto to be $(55.7 \pm 2.5)\%$ [9]. The acceptance of the dimuon trigger hodoscope bank was $(65.3 \pm 1.9)\%$, which includes contributions from geometric acceptance and trigger logic. The uncertainty in the determination of background was estimated to produce at most a 2.0% change in the branching fraction. After calibrating the lead glass there were some residual non-linearities, and this caused the branching fraction to have a small dependence on the photon cluster energy cut. The systematic uncertainty resulting from this cut was estimated to be 0.7%. Taking the branching fraction for $K_L \to \pi^+\pi^-\pi^0$ as $0.1238 \pm 0.0021$ [11] and adding all errors in quadrature the branching fraction is $B(K_L \to \mu^+\mu^-\gamma) = (3.23 \pm 0.23(stat) \pm 0.19(sys)) \times 10^{-7}$. This clearly establishes the decay, while agreeing with the only previous observation of one candidate event [12].

Our result corresponds to $\Gamma_{\mu\mu\gamma}/\Gamma_{\gamma\gamma} = (5.66 \pm 0.59) \times 10^{-4}$. Using Eq. 1 with $f(x) = 1$ gives $\Gamma_{\mu\mu\gamma}/\Gamma_{\gamma\gamma} = B(K_L \to \mu^+\mu^-\gamma)/B(K_L \to \gamma\gamma) = 4.09 \times 10^{-4}$ which is $2.7\sigma$ below our measured value. This is the first indication for the presence of a form factor in the decay $K_L \to \mu^+\mu^-\gamma$. The prediction of Ko [4], $\Gamma_{\mu\mu\gamma}/\Gamma_{\gamma\gamma} = 7.45^{+0.54}_{-0.15} \times 10^{-4}$ differs from our result by $2.9\sigma$.

Assuming the model of Bergström, Massó, and Singer for the form factor, Eq. 1 can be integrated as a function of $\alpha_{K^*}$. This is shown in Fig. 3, using identical expressions for $f_1(x)$ and $f_K(x)$ as those of Ref. [6]. For this model our branching fraction corresponds to $\alpha_{K^*} = -0.018^{+0.131}_{-0.123}$.

The spectrum of dimuon invariant masses is also sensitive to the form factor. In Fig. 4

we show the dimuon mass distribution together with a Monte Carlo prediction for $f(x) = 1$. The background shape, normalized to 10.5 events, was added to the Monte Carlo data. Again the background shape was determined using high-$P_t^2$ events. The assumption that the decay is dominated by the $K\gamma\gamma^*$ vertex leads directly to the dimuon invariant mass spectrum in Eq. 1, and the data in Fig. 4 follows this general shape, apart from the smaller contribution from $f(x)$. Thus the data support the assumption of $K\gamma\gamma^*$ dominance. Also shown in Fig. 4 is the ratio of data to Monte Carlo, i.e. $f(x)^2$. Using a maximum liklihood fit we obtain $\alpha_{K^*} = -0.13^{+0.21}_{-0.35}$ using the dimuon mass spectrum alone. Although less accurate, this is consistent with the value determined from the branching fraction. Together the combined best estimate is $\alpha_{K^*} = -0.028^{+0.115}_{-0.111}$. This differs by $1.8\sigma$ from the world average of $\alpha_{K^*} = -0.28 \pm 0.08$ [11], using $K_L \to e^+e^-\gamma$.

In conclusion, we have measured the branching fraction of the decay $K_L \to \mu^+\mu^-\gamma$ and the result is significantly above the prediction assuming a constant form factor. This supports the BMS model and is $2.9\sigma$ away from the model of Ko. The extracted form factor shows consistency with that from $K_L \to e^+e^-\gamma$, at the $2\sigma$ level. More data, particularly copious $K_L \to e^+e^-\gamma$ with $K_L \to \mu^+\mu^-\gamma$, will better test the viability of the BMS model.

This research was supported in part by the U.S. National Science Foundation and the U.S. Department of Energy. We would like to thank the staff of the Fermilab accelerator, computing and research divisions. One of us (Y.W.W) would like to acknowledge support from an O.J.I. grant from the D.O.E. Another of us (T.N.) would like to thank support from JSPS Fellowships for Japanese Junior Scientists.

# REFERENCES

\* Present address: University of California, Santa Barbara, Santa Barbara, CA 93106.

† Present address: University of Rochester, Rochester, NY 14627.

‡ Present address: Fermi National Accelerator Laboratory, Batavia, IL 60510.

§ Present address: Rutgers University, Piscataway, NJ 08855.

∥ Present address: Colorado State University, Fort Collins, CO 80523.

¶ Present address: Carnegie-Mellon University, Pittsburgh, PA 15213.

\*\* Present address: University of West Virginia, Morgantown, WV 26506.

†† Present address: Center for Neighborhood Technology, 2125 West North Ave., Chicago, IL 60647.

‡‡ Present address: Seoul National University, Seoul 151-742,Korea.

[1] L. Bergström, E. Massó, P. Singer, Phys. Lett. **131**B, 229 (1983).

[2] P. Singer, Nuc. Phys. A**527**, 713c (1991).

[3] G. Belánger, C. Q. Queng, Phys. Rev. D **43**, 140 (1991).

[4] P. Ko, Phys. Rev. D **44**, 139 (1991).

[5] G. D. Barr *et al.*, Phys. Lett. **240**B, 283 (1990).

[6] K. E. Ohl *et al.*, Phys. Rev. Lett. **65**, 1407 (1990).

[7] L. Littenberg and G. Valencia, Ann. Rev. Nuc. Part. Phys. **43**, 729 (1993).

[8] D. A. Harris *et al.*, Phys. Rev. Lett. **71**, 3914 (1993), and D. A. Harris *et al.* Phys. Rev. Lett. **71**, 3918 (1993).

[9] M. B. Spencer, UCLA Ph.D. thesis, unpublished (1995).

[10] K. S. McFarland *et al.*, Phys. Rev. Lett. **71**, 35 (1993).

[11] Particle Data Group, L. Montanet *et al.*, Phys. Rev. D **50**, 1173 (1994).

[12] A. S. Carroll *et al.*, Phys. Rev. Lett. **44**, 525 (1980).

FIGURES

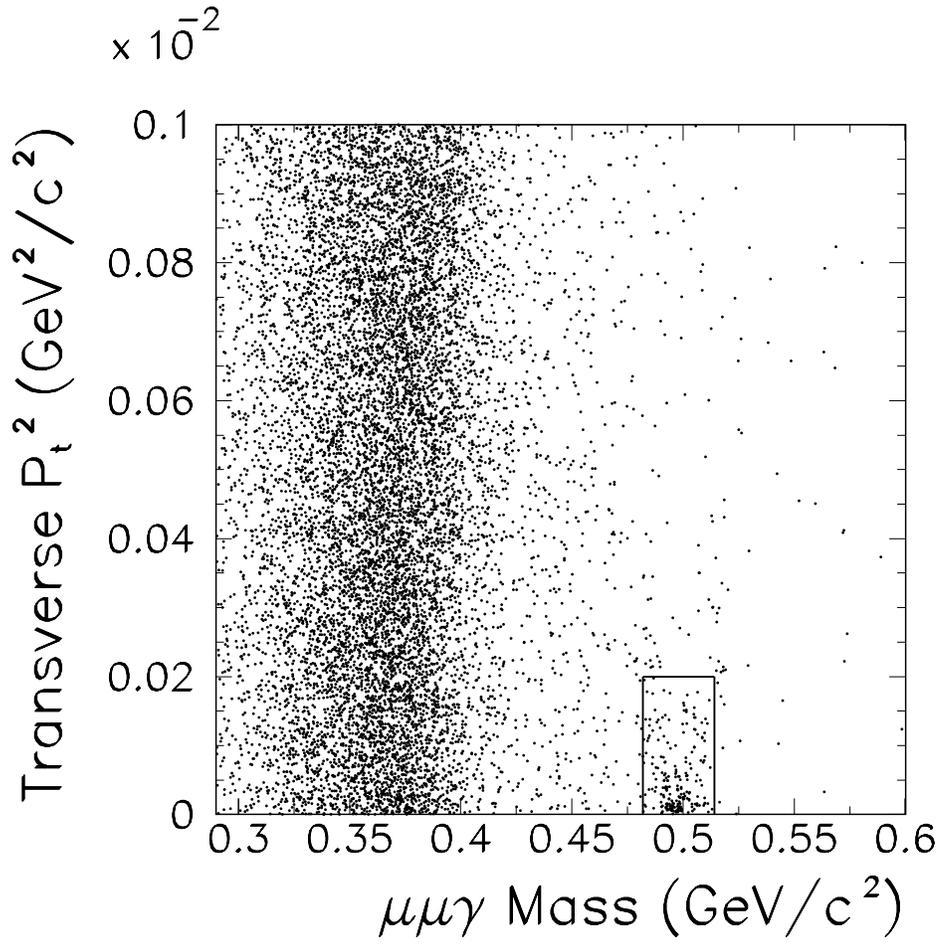

FIG. 1. $P_t^2 - vs - \mu^+\mu^-\gamma$ invariant mass distribution for events after cuts.

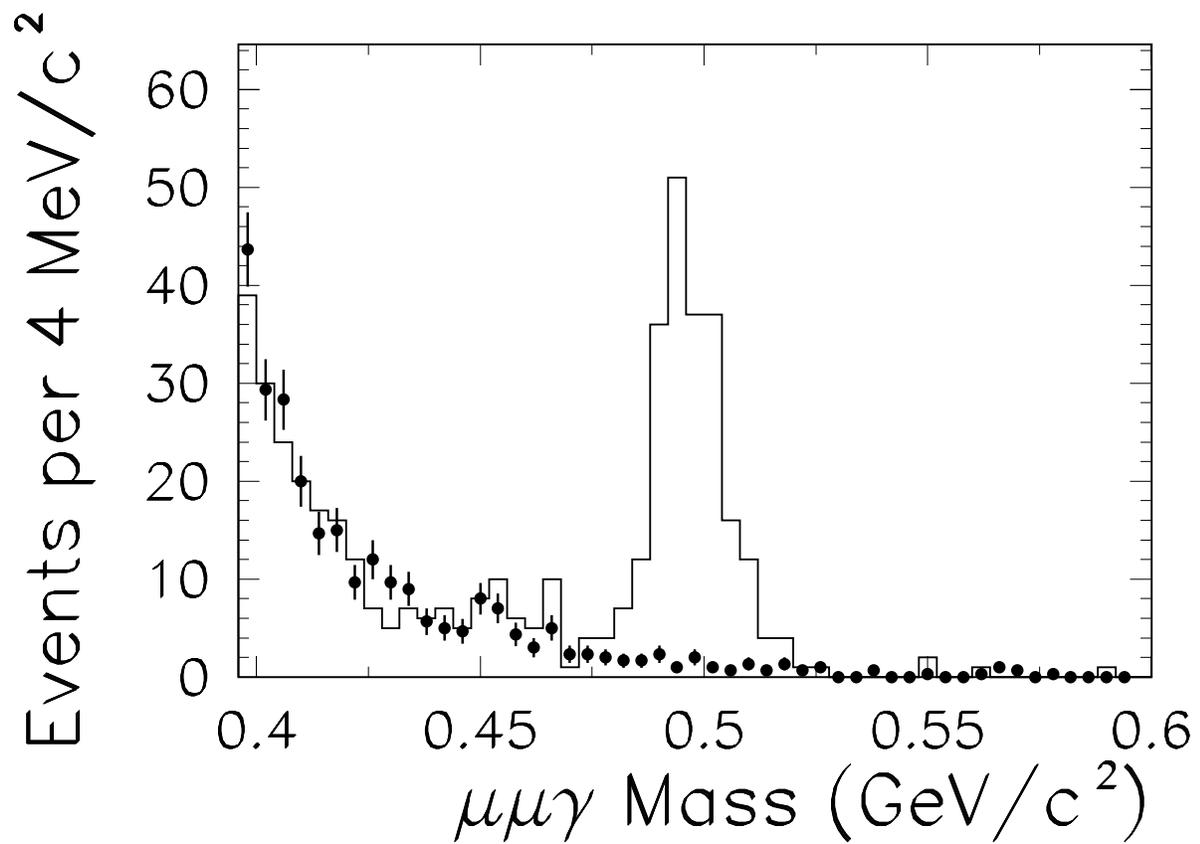

FIG. 2. Distribution of $\mu^+\mu^-\gamma$ invariant mass after cuts. Histogram is data for events with $P_t^2 < 200 MeV^2/c^2$ and points are data for events with $400 < P_t^2 < 1000 MeV^2/c^2$, and normalized to the low $P_t^2$ data.

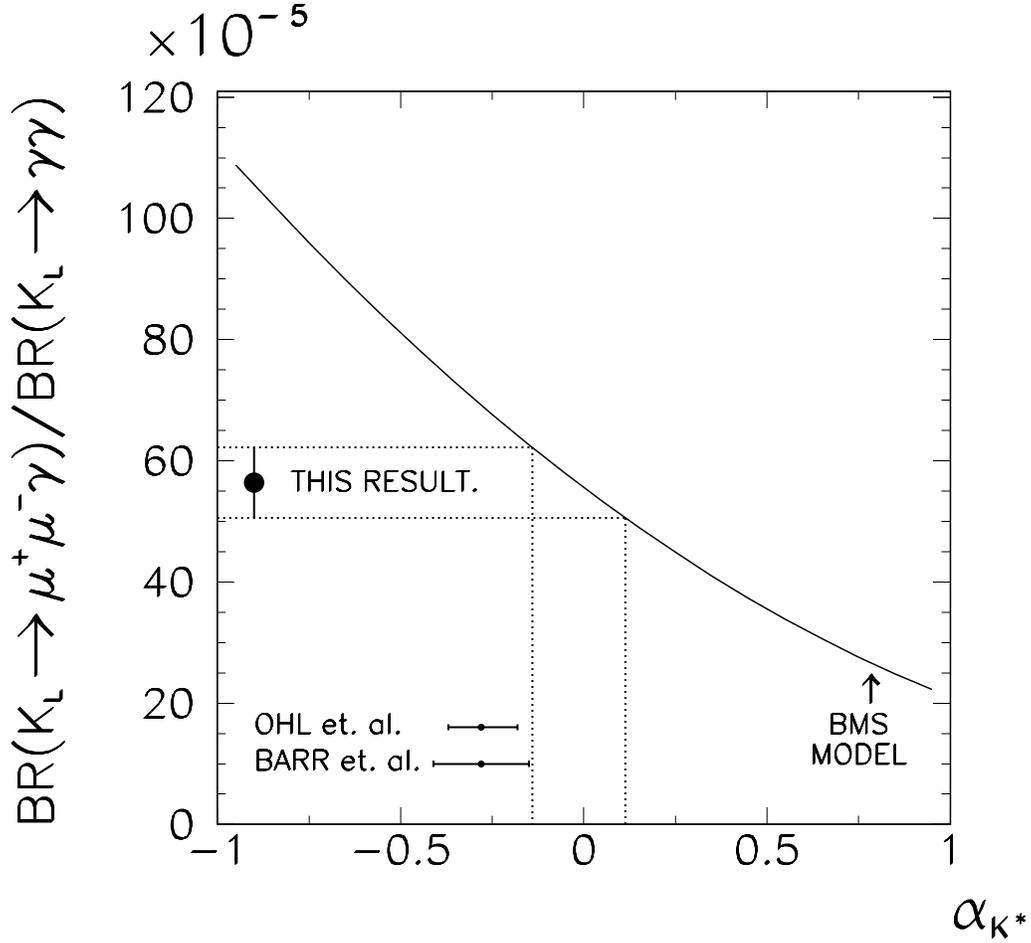

FIG. 3. This result for the branching ratio of $K_L \to \mu^+\mu^-\gamma$, normalized to $K_L \to \gamma\gamma$ (large black dot) and the extracted range of the parameter $\alpha_{K^*}$ from the model of Bergström, Massó, Singer [1]. Also shown (horizontal error bars) are the extracted values of $\alpha_{K^*}$ from Refs. [6] and [5] for the case of $K_L \to e^+e^-\gamma$.

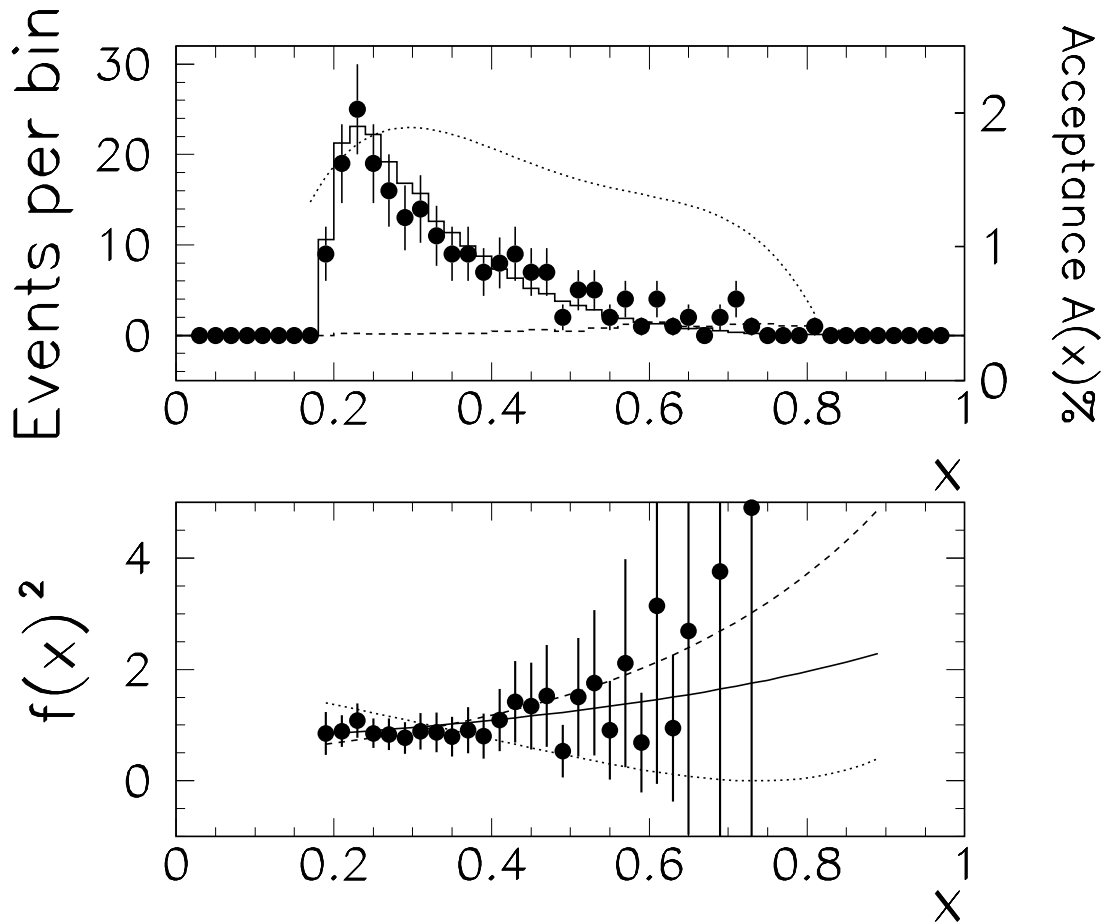

FIG. 4. Top: Dimuon mass distribution, points are data, histogram is a Monte Carlo simulation with $f(x) = 1$, dashed line is the estimated background, dotted line is the acceptance. Bottom: the ratio of data to Monte Carlo from the top plot (*i.e.* our measured values of $f(x)^2$). The solid line is the result from a maximum liklihood fit. For illustration we also show predictions assuming $\alpha_{K^*} = -1$ (dashed) and $\alpha_{K^*} = +1$ (dotted). For these plots the normalization was fixed by the number of events.

14